\documentclass[11pt,twoside]{article}

\usepackage{asp2006} \usepackage{epsf} \usepackage{lscape}
\usepackage{graphicx}

\begin{document}
\title{Hubble Space Telescope Imaging of the Expanding Nebular Remnant
of the 2006 Outburst of RS Ophiuchi}
 
\author{D.~J.~Harman,$^{1}$ M.~F.~Bode,$^{1}$ M.~J.~Darnley,$^{1}$ T.~J.~O'Brien,$^{2}$
H.~E.~Bond,$^{3}$ S.~Starrfield,$^{4}$ A.~Evans,$^{5}$ S.~P.~S.~Eyres$^{6},$
V.~A.~R.~M.~Ribeiro,$^{1}$ and J.~M.~Echevarria$^{7}$}
\affil{$^{1}$~Astrophysics Research Institute, Liverpool John Moores
  Univ, UK;\\ $^{2}$~Jodrell Bank Observatory, University of
  Manchester, UK;\\ $^{3}$~Space Telescope Science Institute, Baltimore,
  MD, USA;\\ $^{4}$~Arizona State University, Tempe, AZ, USA;\\
  $^{5}$~Astronomy Group, Keele University, UK;\\ $^{6}$~Centre for Astrophysics,
   University of Central Lancashire, UK;\\ $^{7}$~Instituto de Astronomia, Universidad Nacional Autonoma de Mexico, A.P. 70-264, Mexico}


\begin{abstract}
We report {\it Hubble Space Telescope} imaging obtained 155 days and
449 days after the 2006 outburst of RS Ophiuchi.  Both epochs show
evidence of extended emission, consistent with that seen in earlier
radio observations, and a maximum expansion rate of
$3200\pm300$ km s$^{-1}$ (in the plane of the sky).  The extended structure is consistent with
the remnant having a bipolar morphology with an inclination similar to
that determined for the binary.
\end{abstract}

\section {Introduction}

On 12.83 February 2006 the recurrent nova RS Ophiuchi was observed to
be undergoing its sixth observed eruption \citep[][$t=0$]{Narumi2006_hst}.
VLBI imaging at 6 and 18 cm began 13.8 days after outburst
\citep{2006Natur.442..279_hst}.  These initial images showed a partial
ring of non-thermal (synchrotron) emission, of radius 13.8 AU (assuming a
distance of 1.6 kpc), consistent with emission from the forward shock.  Extended lobes aligned in the east-west direction then gradually emerged,
with the eastern lobe appearing first.  This morphology is consistent
with that derived from VLBI observations 77 days after the 1985
outburst \citep{1989MNRAS.237...81T_hst}.  O'Brien et al. (2006)
proposed a simple geometrical model comprising an expanding
double-lobed structure, with the major axis perpendicular to the
proposed plane of the central binary orbit.

Here we briefly report {\it Hubble Space Telescope} ({\it HST}) observations of the
expanding nebular remnant taken at $t=155$ days (Epoch 1) and $t=449$ days (Epoch 2),
comparing both epochs to the structures seen earlier in the radio.
Bode et al. (2007, hereafter Paper I) provide a much fuller
discussion of the Epoch 1 {\it HST} observations. 

\section {HST Observations and Data Reduction}

RS Oph was observed (prop. ID 11004) in Director's Discretionary Time (DDT)
155 days after outburst on 17th July 2006 using the Advanced Camera
for Surveys (ACS) on the {\it HST}.  The 0.025\arcsec pixel$^{-1}$\
High-Resolution Channel (HRC) was used together with three narrow-band filters
to isolate the H$\alpha$+N\,[{sc ii}] (filter F658N), [O\,{\sc iii}]\ $\lambda$5007\,\AA\
(F502N) and [Ne\,{\sc v}]\ $\lambda$3426\,\AA\ (F343N) nebular emission lines.  One orbit was used to image RS Oph and the second to observe HD 166215, a bright nearby star of similar
spectral type to be used as a PSF standard.

The Epoch 2 observations also took place in DDT, this time at 449 days after outburst on 7th May 2007, using the
Wide Field Planetary Camera 2 (WFPC2, utilised due to the failure of the ACS).  Three orbits of DDT were used to image RS~Oph through the
[O\,{\sc iii}]\ $\lambda$5007\,\AA\ filter alone.

All data were processed using standard procedures outlined in the {\it HST} ACS
Data Handbook and the Pydrizzle and Multidrizzle Handbooks, using
optimal input parameters to maximise the signal-to-noise of each
image.  Using the profile of HD 166215 as the PSF, as well as profiles generated
by TinyTim \citep{Krist2005_hst}, deconvolution using the Lucy-Richardson method was
performed on each of the RS Oph images.  Tests using both CLEAN and
Maximum Entropy techniques produces similar results (see Paper I for
a more detailed discussion of data reduction methodology).

As part of this study, a re-analysis was carried out of pre-outburst
WFPC2 observations in the [O\,{\sc iii}]\ $\lambda$5007\,\AA\  line on 12th June 2000.  No
extended emission was detectable, confirming the results of
Brocksopp et al. (2003).

\section {Results and Discussion}

As can be seen in Figure~1, extended structure was clearly visible in
the Epoch 1 [O\,{\sc iii}]\ $\lambda$5007\,\AA\ PSF-subtracted image
(and was indeed visible in
the raw image).  Deconvolution revealed more detailed structure in
both [O\,{\sc iii}]\ $\lambda$5007\,\AA\ and [Ne\,{\sc v}]\ $\lambda$3426\,\AA.  There was also a hint of
possible extended
emission close to the central source in the H$\alpha$ line, but this was
not present at a significant level.

\begin{figure}[!ht]
\includegraphics[width=\textwidth]{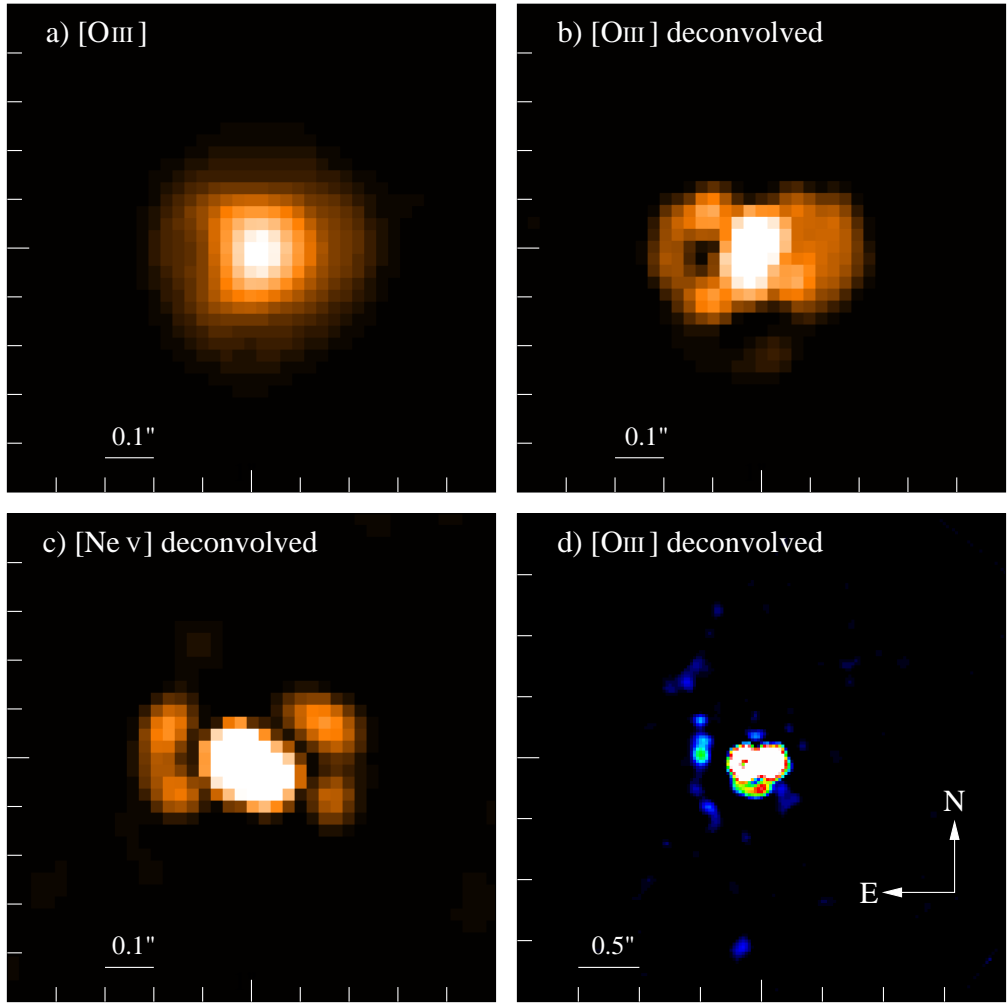}
\caption{{\it HST} ACS/HRC narrow-band Epoch 1 images of RS Oph. (a)
  PSF-subtracted [O\,{\sc iii}]\ $\lambda$5007\,\AA\ image clearly showing extended emission at
  subarcsecond size scales particularly in the east-west direction; (b)
  results of deconvolution of the [O\,{\sc iii}]\ $\lambda$5007\,\AA\ RS Oph PSF-subtracted image with the
  PSF star, showing a double ring structure; (c) same as (b), but for
  [Ne\,{\sc v}]\ $\lambda$3426\,\AA\ and using a TinyTim PSF; (d) deeper, larger area view of (b),
  showing an arc-like feature to the east and a southern blob of emission.}
\end{figure}

In the deconvolved [O\,{\sc iii}] and [Ne\,{\sc v}] Epoch 1 images,
the most striking feature
is an apparent double ring structure with major axis lying east-west and total
(peak-to-peak) extent $360\pm30$ mas ($580\pm50$ AU at $d=1.6$ kpc).
The most extended radio structures \citep[the outer
  lobes;][]{2006Natur.442..279_hst} lie along this axis.  Assuming
ejection at $t=0$ the expansion rate of the optically emitting gas along
this axis is $1.2\pm0.1$ mas day$^{-1}$ (equivalent to
$v_{\mathrm{exp}}=3200\pm300$ km s$^{-1}$ in the plane of the sky).
The optical emission is also detectable above background in the
deconvolved images to a total extent of $520\pm50$ mas, corresponding
to an expansion rate of $1.7\pm0.2$ mas day$^{-1}$.  We compare this
to $1.4\pm0.3$ mas day$^{-1}$ for the east-west lobes seen in the
radio, taken over four epochs from $t=21.5$ to 62.7 days during the
2006 outburst (O'Brien et al., in preparation) and 1.3 mas day$^{-1}$
for the equivalent features derived from VLBI observations during the
1985 outburst \citep{1989MNRAS.237...81T_hst}.  Thus, there is evidence
that the bipolar emission seen here and in the radio arises from the
same regions of the remnant, if the expansion velocities are roughly
constant.

O'Brien et al. 2006) proposed a simple bipolar model for the
radio emission, where optical depth evolution led to a gradual
``uncovering'' of various features, most notably the outermost radio
lobes.  We have modeled the optical emission seen in our Epoch 1 {\it
  HST} images with a ``peanut-shaped'' bipolar structure using the
code described by Harman \& O'Brien (2003, see their Fig. 2).  As can be
seen in Figure 2 of Paper I, the model reproduces the morphology of the
optical emission extremely well for an inclination $i=35\deg$,
consistent with the major axis of the optical nebula lying normal to the binary plane.  For
$d=1.6\pm0.3$ kpc \citep{Bode1987_hst} and such an inclination, the true
expansion velocity, $v=5600\pm1100$, and $v_{\mathrm{rad}}=4600\pm900$
km s$^{-1}$ are implied for the material at the poles.  Unless the
ejection of material in the explosion is highly anisotropic, this
result reinforces the notion that remnant shaping occurs due to the
interaction of the ejecta with the pre-existing circumstellar environment
\citep[see][]{2008arXiv0804.2628W_hst}.

The Epoch 2 data are shown in Figure~2, and again
extended structure is clearly visible in the deconvolved image.  The peak-to-peak extent along the east-west axis
after 449 days was 1.1\arcsec ($1,770$ AU at $d=1.6$ kpc).  This
translates to an expansion rate along the east-west axis of 1.2 mas day$^{-1}$.

\begin{figure}[ht]
\includegraphics[width=\textwidth]{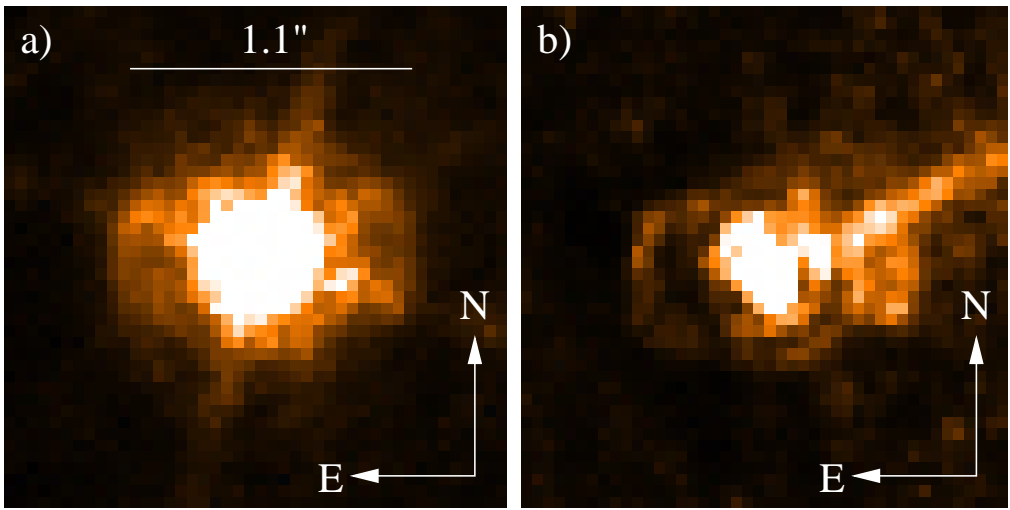}
\caption{{\it HST} WFPC2 narrow-band Epoch 2 images of RS Oph. (a)
  ``Raw'' [O\,{\sc iii}]\ $\lambda$5007\,\AA\ image; (b) results of deconvolution of
  the raw [O\,{\sc iii}] image with a PSF star; a double ring structure is again visible.  The NW spike is an
artefact enhanced by image deconvolution.}
\end{figure}

It is clear that the
current spherically symmetric hydrodynamic models used to describe
the shock evolution in RS Oph need revision to account for the
geometry revealed by these observations and those in the radio.
Work currently underway includes the full analysis of the Epoch 2 data in conjunction with optical spectroscopy
to explore more details of the remnant evolution.

\acknowledgements

The authors are very grateful to the {\it HST} Director for provision
of Discretionary Time and the {\it HST} support staff for their
assistance with planning and analysing the observations reported
here.

\end{document}